\documentclass[amsmath,amssymb,nofootinbib,showpacs,twocolumn,nofootinbib]{revtex4-1}
\usepackage{dcolumn}
\usepackage{bm}
\usepackage{graphicx}
\usepackage{epstopdf}
\usepackage{epsfig}
\usepackage{color}
\usepackage{wasysym}
\usepackage{natbib}
\usepackage{twoopt}
\usepackage{dashrule}
\usepackage[breaklinks=true,colorlinks=true,linkcolor=blue,citecolor=blue,urlcolor=blue,pdfauthor={Tomassetti},pdftitle={Wanted}]{hyperref}
\def\MyTitle#1{{\textit{#1}}}
\def\Journal#1#2#3#4{{#4}, {#1}, {#2}, #3} 

\newcommand{\ApJ}{ApJ}
\newcommand{\AeA}{A\&A}
\newcommand{\PRL}{PRL}
\newcommand{\PRD}{PRD}
\newcommand{\etal}{et al.}

\newcommand{\AMS}{\textsf{AMS}}
\newcommand{\Fermi}{\textit{Fermi}-LAT}
 
\newcommand{\ie}{\textit{i.e.}} 

\newcommand{\Hyd}{\textsf{H}}
\newcommand{\p}{\textsf{p}}
\newcommand{\He}{\textsf{He}}
\newcommand{\Li}{\textsf{Li}}
\newcommand{\Be}{\textsf{Be}}
\newcommand{\B}{\textsf{B}}
\newcommand{\C}{\textsf{C}}
\newcommand{\N}{\textsf{N}}
\newcommand{\Oxy}{\textsf{O}}
\newcommand{\Fe}{\textsf{Fe}}
\newcommand{\BC}{\textsf{B}/\textsf{C}}
\newcommand{\eplus}{\ensuremath{e^{+}}}
\newcommand{\pfrac}{e\ensuremath{^{+}}/(e\ensuremath{^{-}}\,+\,e\ensuremath{^{+}})} 
\newcommand{\pbarp}{\ensuremath{\bar{p}/p}}

\newcommand{\CO}{\textsf{C/O}}
\newcommand{\OFe}{\textsf{O/Fe}}
\newcommand{\CFe}{\textsf{C/Fe}}
\newcommand{\NeFe}{\textsf{Ne/Fe}}
\newcommand{\MgFe}{\textsf{Mg/Fe}}
\newcommand{\ArFe}{\textsf{Ar/Fe}}
\newcommand{\CaFe}{\textsf{Ca/Fe}}

\begin{document}
\title{Origin of the spectral upturn in the cosmic-ray \CFe{} and \OFe{} ratios}
\author{Nicola Tomassetti}
\address{LPSC, Universit\'e Grenoble-Alpes, CNRS/IN2P3, F-38026 Grenoble, France; email: nicola.tomassetti@lpsc.in2p3.fr}
\date{September 2015} 
%%%%%%%%%%%%%%%%%%%%%%%%%%%%%%%%%%%%%%%%%%%%%%%%%%%%%%%%%%%%%%%%%%%%%%%%%%%%%%%%%%%%%%%%%%%%%%%%%%%%%%%%%%%%%%%%%%%%%%%%%
%
%%%%%%%%%%%%%%%%%%%%
\begin{abstract} %%%
%%%%%%%%%%%%%%%%%%%%
The observed spectrum of Galactic cosmic rays has several exciting features such as the rise in the positron fraction 
above $\sim$\,10\,GeV of energy and the spectral hardening of protons and helium at $\gtrsim$\,300\,GeV/nucleon of energy. 
The ATIC-2 experiment has recently reported an unexpected spectral upturn in the elemental ratios involving iron, such as 
the \CFe{} or \OFe{} ratios, at energy $\gtrsim$\,50\,GeV per nucleon. It is recognized that the observed positron excess 
can be explained by pion production processes during diffusive shock acceleration of cosmic-ray hadrons in nearby sources.
Recently, it was suggested that a scenario with nearby source dominating the GeV-TeV spectrum may be connected with the 
change of slope observed in protons and nuclei, which would be interpreted as a flux transition between the local component 
and the large-scale distribution of Galactic sources.
Here I show that, under a two-component scenario with nearby source, the shape of the spectral transition is expected to be 
slightly different for heavy nuclei, such as iron, because their propagation range is spatially limited by inelastic 
collisions with the interstellar matter. This enables a prediction for the primary/primary ratios between light and heavy 
nuclei. From this effect, a spectral upturn is predicted in the \CFe{} and \OFe{} ratios in good accordance with the ATIC-2 data.
\end{abstract}
 % nuclear astrophysics: spallation reaction

\maketitle

%%%%%%%%%%%%%%%%%%%%%%%%%%%%%%
\MyTitle{Introduction} --- %%%
%%%%%%%%%%%%%%%%%%%%%%%%%%%%%%
%
The observed cosmic-ray (CR) spectrum has several unexplained features
such as the 10--200\,GeV rise of the positron fraction \pfrac{} \citep{Adriani2009,Serpico2011}
and the spectral hardening of proton and helium above $\sim$\,300\,GeV/nucleon of energy \citep{Adriani2011,Yoon2010,Blasi2013},
that are now being investigated with high precision by the \AMS{} experiment \citep{Aguilar2015,Accardo2014}.
Recently, a puzzling spectral upturn has been reported by the ATIC-2 experiment for nuclear ratios involving
iron, such as the \CFe{} or \OFe{} ratios at $\sim$\,50\,GeV/nucleon of energy \citep{Panov2013,Panov2014}. 
In the traditional descriptions, \emph{primary} CRs such as electrons, protons, \He, \C-\N-\Oxy, or \Fe{} nuclei 
are injected in the interstellar medium (ISM)
by a continuous distribution of Galactic sources, 
after being accelerated to power-law spectra $\sim\,E^{-\nu}$, with $\nu\approx$\,2--2.4, up to PeV energies.
Their spectrum is steepened by diffusive propagation in the Galactic halo (typical half-size $L\sim$\,3-10\,kpc)
with diffusion coefficient $K\propto E^{\delta}$, where $\delta\sim$\,0.3--0.7.
Interactions of CRs with this gas of the Galactic disk (half-size $h\sim$\,100\,pc) give rise 
to \emph{secondary} particles such as \eplus{} or \Li-\Be-\B{} nuclei, that are expected to be
$E^{\delta}$-times steeper than primary CRs.
The several models based on this picture agree in
predicting smooth power-law spectra for primary nuclei, almost energy-independent primary/primary ratios,
and a positron fraction decreasing steadily as $\sim E^{-\delta}$ \cite{Grenier2015}. 
The recently observed features in CR leptons, protons, and heavier nuclei are clearly at tension with these predictions.
The observed positron excess requires an additional leptonic component that may come
from nearby exotic sources, such as dark-matter particles annihilation, 
or known sources, such as pulsars or \emph{old} supernova remnants (SNRs) \citep{Serpico2011}.  
In the \emph{old SNR} scenario, the excess is produced by interactions of CR protons undergoing 
acceleration in proximity of the shock waves \citep{Blasi2009}. The $e^{\pm}$ production and subsequent re-acceleration 
gives rise to a SNR component which is \emph{harder} than that of primary protons or electrons, $E^{-\nu}$,
and may explain the \AMS{} data \citep{MertschSarkar2014}.
Other secondary species, such as \Li-\Be-\B{} nuclei or antiprotons, are also expected to be 
produced in a similar way.
Assuming that the observed CR flux is entirely provided by this type of sources,
this mechanism predicts a \emph{rise} of the \BC{} ratio at $\sim$\,100\,GeV 
per nucleon \citep{MertschSarkar2009}.
However the measured \BC{} ratio does not show such a feature \citep{CholisHooper2014}.

\begin{figure}[!t] 
  \includegraphics[width=0.42\textwidth]{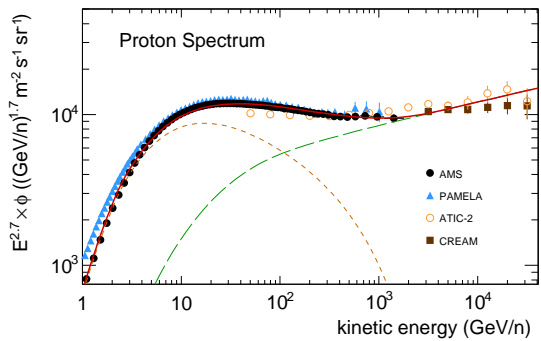}
\caption{\footnotesize%
  Energy spectrum of CR protons multiplied by $E^{2.7}$.
  The solid lines indicate the model calculations. The contribution arising from
  the nearby SNR (short-dashed lines) and from the Galactic SNR ensemble (long-dashed lines) are shown.
  The data are from \AMS{} \citep{Aguilar2015}, PAMELA \citep{Adriani2011}, ATIC-2 \citep{Panov2009}, and CREAM \citep{Yoon2010}. 
}\label{Fig::ccProtonSpectrum}
\end{figure}

In \citet{TomassettiDonato2015}, we have shown that the \emph{old} SNR scenario is  incomplete in order 
to account for the observations of CR hadronic spectra at TeV--PeV energies, 
because these energies can be only attained with a magnetic field amplification mechanism which,
in turn, is not compatible with secondary production at the shock \citep{Serpico2011,Kachelriess2011}. 
Besides, the spectral hardening of CR proton and helium suggests that
different types of sources may contribute to their flux \citep{Vladimirov2012}.
In our two-component scenario, the total CR flux is described by 
a \emph{nearby} source component $\phi^{L}$ in the $\sim$\,GeV--TeV region, 
arising from an \emph{old} SNR, 
and by a \emph{Galactic ensemble} SNR component $\phi^{G}$, 
arising from younger sources with amplified magnetic fields, in the $\sim$\,TeV--PeV region. 
A key consideration is that, due to Compton and synchrotron losses, the $e^{\pm}$ propagation length 
is limited within a typical distance $\lambda^{\rm rad} \sim \sqrt{\tau^{\rm rad} K} \propto E^{(\delta-1)/2}$
with cooling time $\tau^{\rm rad}\sim 300\times E^{-1}$\,Myr\,GeV$^{-1}$.
A nearby source (within a few 100\,pc) seems, therefore, necessary to explain the GeV-TeV $e^{\pm}$ flux \citep{Delahaye2010}.
In contrast, CR protons and light nuclei do not experience radiative losses, so that their local flux may 
arise from the contribution of a larger population of Galactic sources \citep{Taillet2003}.
As shown, such a scenario may account both for the rise in the positron fraction and for the decreasing of the \BC{} ratio.
The new \AMS{} proton data, shown in Fig.\,\ref{Fig::ccProtonSpectrum}, are also well consistent with a smooth
flux transition as described by model.
%%%%%%%%%%%%% PROTON SPECTRUM %%%%%%%%%%%% %%%%%%%%%%%%%%%%%%%%%%%%%
%%%%%%%%%%%%%%%%%%%%%%%%%%%%%%%%%%%%%%%%%%%%%%%%%%%%%%%%%%%%%%%%%%%%%

In this paper I show that, under such a scenario with nearby source,
the shape of the spectral transition between the two compontents
has a characteristic signature in the spectrum of heavy nuclei, 
due to a combination of propagation and spallation effects.
In fact, the propagation range of heavy nuclei like iron is spatially limited 
by inelastic collisions with the ISM nuclei, 
which may prevent CRs injected from distant sources to reach the Solar System.
To study this effect, I make use of an effective calculation scheme, based on the propagation scale length, 
that enables a prediction for the ratios between light and heavy primary nuclei such as \CFe{} and \OFe. 
For these ratios, a remarkable spectral upturn is predicted at $\sim$\,50\,GeV of energy.

%%%%%%%%%%%%%%%%%%%%%%%%%%%%%%
\MyTitle{Calculations} --- %%%
%%%%%%%%%%%%%%%%%%%%%%%%%%%%%%
%
In conventional calculations based on the diffusion approximation, as long as all sources have 
the same spectral properties, the model predictions for the spectra of CR nuclei at Earth are known 
to be only barely sensitive to the exact distribution of Galactic sources \citep{Genolini2015}.
But in the case of distinct classes of sources characterized by different properties,
it becomes important to account for the SNR spatial distribution \citep{Taillet2003}.
This is indeed the case for the propagation of heavy nuclei in our two-component scenario,
where the total observed flux arise from the superposition of two classes of SNRs, $S^{\rm L}$ and $S^{\rm G}$, 
that inject CRs in the ISM with different spectral shape. 
In particular, the Galactic ensemble component, $S^{G}$, reflects the contribution 
of a large-scale SNR population that extend to several kpc of distance.
Thus, the fraction of these SNRs effectively contributing to the local observed flux
depends on the propagation properties of the considered element.
To first approximation, the typical propagation scale distance of CR nuclei, $\lambda^{\rm sp}\equiv\sqrt{K\tau^{\rm sp}}$, 
can be estimated by using a spallation time-scale $\tau^{\rm sp}\cong \frac{L}{h\Gamma^{\rm sp}}$.
Here, to effectively account that CRs interact only where they cross the disk, the matter density 
is considered as \emph{diluted} in the propagation region by the $h/L$ ratio \citep{Jones1978}.   
In contrast to leptons, the function $\lambda^{\rm sp}$ for CR nuclei \emph{increases} with energy and 
\emph{decreases} with the mass. Roughly, the interaction cross sections increase with the projectile 
mass as $\sigma^{\rm sp} \propto M^{0.7}$ \citep{Letaw1983}, giving $\lambda^{\rm sp}(E) \propto M^{-0.35}E^{\delta/2}$. 
This trend illustrates that, for the CR spectrum detected at Earth, 
heavier nuclei must come from sources located in nearer regions.
Clearly, this reduces the fraction of Galactic sources that effectively contribute observed flux at Earth. 
In order to estimate this fraction for the relevant CR species, one has to model a realistic
distribution for the Galactic SNRs as function of the distance to the Earth.  
For this purpose, I follow closely the effective approach of \citet{Ahlers2009}, where the spatial distribution of 
SNRs is determined by a toy Monte Carlo generation of randomly distributed sources drawn from a probability density function.
The input function assumes a four-armed Galactic structure, which has been confirmed by a recent analysis \citep{Camargo2015}. 
The calculation provides the SNR distribution function in terms of Earth centered coordinates and integrated over the polar coordinate. 
%
%%%%%%%%%%%%%%%%%%%%%%%%%%%%%%%%%%%%%%%%
\begin{figure}[!t] 
  \includegraphics[width=0.45\textwidth]{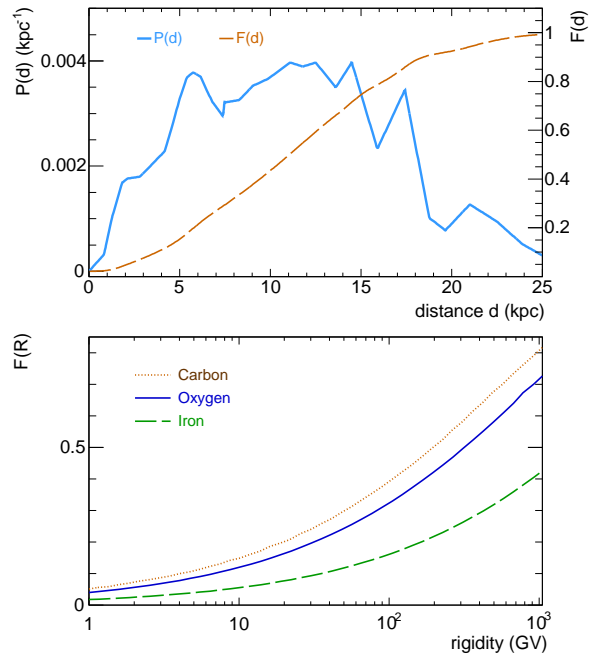}
\caption{\footnotesize%
Top: Distribution function (solid line) and cumulative function (dashed line) of Galactic SNRs as function of the distance $d$ from the solar system.
Bottom: Fraction of Galactic SNRs contributing to the CR flux of \Oxy{} (solid line) and \Fe{} (dashed line).
}\label{Fig::ccSNRDensity}
\end{figure}
%%%%%%%%%%%%%%%%%%%%%%%%%%%%%%%%%%%%%%%%
The normalized probability density of SNRs, $P(d)$, is shown in Fig.\,\ref{Fig::ccSNRDensity} (top) as a function of the distance $d$ from the Earth.
As seen, the contribution is probabilistically suppressed within $\sim$\,1.5\,kpc due to the interarm position of the Solar System;
but this does not prevent the factual occurrence of one (or few) anomalous SNR event in the Solar neighborhood
that would manifest itself as a distinctive component of the local CR spectrum. 
In the figure, it is also shown the cumulative fraction of SNRs falling within a certain distance $d$, 
$F(d)=\int_{0}^{d}P(l) dl$.
From this information, the fraction of Galactic SNRs contributing to the CR flux detected for a $j$-type element is estimated as
$F(\lambda^{\rm sp}_{j})$, where $\lambda^{\rm sp}_{j}$ can be expressed as function of energy or rigidity.

The model setup follows closely our earlier work \citep{TomassettiDonato2015,TomassettiDonato2012}. I briefly outline the key parameters. 
For the local SNR component, the magnetic field is $B=$\,1\,$\mu$G and the upstream fluid speed is $u_{1}=$\,5$\times$\,10$^{7}$\,cm\,s$^{-1}$.
The SNR age is $\tau^{\rm snr}=$\,50\,kyr. Its maximum rigidity is $R^{\rm max}=$\,1\,TV. 
A damping factor $\kappa_{B}=$\,16 is used to enhance the (otherwise) Bohm-like diffusivity at the shock.
These properties are typical for SNRs at their late evolutionary stages. 
For the large-scale population of the Galactic ensemble, represented by younger SNRs with strong shocks and amplified magnetic fields,
typical parameters are $u_{1}\sim$\,10$^{9}$\,cm\,s$^{-1}$, $B/\kappa_{B}\sim$\,100\,$\mu$G, and $R^{\rm max}\sim$\,5\,PV. 
No secondary production occurs in SNRs with these properties, and their spectrum, $S^{G}\sim\,R^{-\nu}$,
is independent on the exact values of the environmental parameters.
The spectral indices are taken as $\nu=$\,2.2 and 2.1 for $Z=1$ and $Z>1$, respectively,
while the spectra from the \emph{old} SNR component are softer by 0.5 for all elements.
The spectral indices of the SNR ensemble agree with the basic DSA predictions and with $\gamma$--ray 
observations of young SNRs \citep{Blasi2013}. 
On the contrary, softer spectra may arise from weak shocks or from environmental effects such as interaction between 
shock and dense gas or turbulence damping, which may well be the case in \emph{old} SNRs.
The elemental dependence of the spectral indices is a known feature of the CR spectrum, 
possibly ascribed to a $M/Z$--dependent injection efficiency in SNR shocks \citep{Malkov2012}.
The source abundances and the cross sections for destruction/production processes 
are those adopted from previous studies \citep{Tomassetti2012,TomassettiDonato2012}. 
The diffusion coefficient is taken as a universal function in rigidity, $K(R) = \beta K_{0}(R/R_{0})^{\delta}$, 
with $K_{0}/L=$\,0.1/5\,kpc\,Myr$^{-1}$ and index $\delta=$\,1/2, as expected from an 
Iroshnikov-Kraichnan turbulence spectrum and tested to the new \BC{} data from PAMELA.
The ISM is assumed to be composed by 90\,\% \Hyd{} and 10\,\% \He, with
surface density $2h\times n^{\rm ism}$=\,200\,pc $\times$\,1\,cm$^{-3}$.
The solar modulation is described under the \textit{force-field} approximation \citep{Gleeson1968}.
The proton spectrum is shown in Fig.\,\ref{Fig::ccProtonSpectrum}. 
using a modulation potential $\Phi=800$\,MV to describe the new \AMS{} data. 
The spectrum is described a superposition of two source components $\phi_{p}=\phi_{p}^{L} + \phi_{p}^{G}$, 
shown as dashed lines, that amounts to  85\,\% for the nearby SNR and 15\,\% for the ensemble, at 1\,GeV/n.

To account for the propagation/spallation effect described above, the source term for the Galactic SNR component, $S_{j}^{G}(E)$,
is replaced by the \emph{effective source term} $\hat{S}_{j}^{G}\equiv F_{j}(E)\times S_{j}^{G}(E)$ for all $j-$types nuclei, 
where $F_{j}(E)\equiv F(\lambda^{\rm sp}_{j}(E))$, is the fraction of Galactic SNRs contributing the nuclear species $j$-th.
In the high-energy limit one has $\hat{S}^{G}\rightarrow S^{G}$ for all species, but lighter CR nuclei 
experience a more rapid convergence than heavier nuclei, due to interactions.
Typical cross sections for collisions with the ISM are $\sigma^{\rm sp} \sim$\,40\,mb for protons, 
$\sim$\,300\,mb for \C-\N-\Oxy{}, and $\sim$\,900\,mb for \Fe.
The function $F(R)$ is shown Fig.\,\ref{Fig::ccSNRDensity} (bottom) as function of rigidity for \C, \Oxy{} and \Fe. 
%%%%%%%%%%%%% HEAVY NUCLEI SPECTRA - O & Fe %%%%%%%%%%%%%%%%%%%%%%%%%
\begin{figure}[!t] 
  \includegraphics[width=0.45\textwidth]{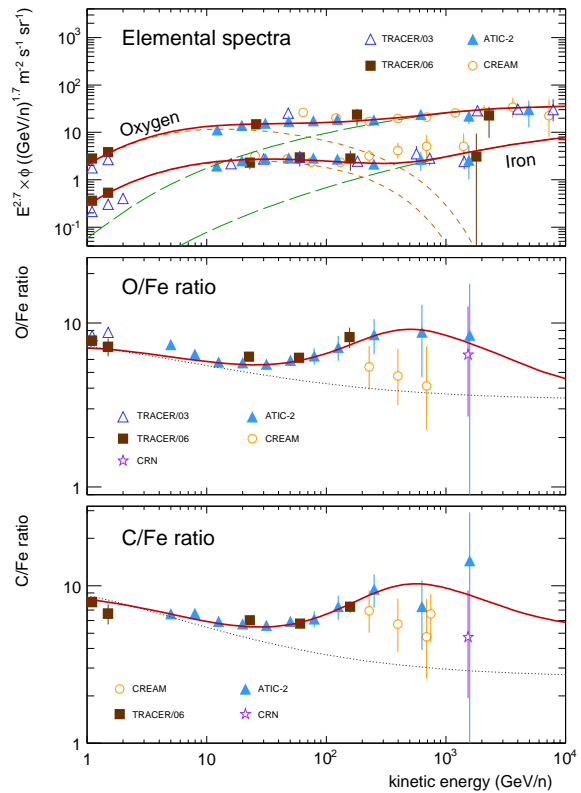}
\caption{\footnotesize%
  Energy spectra of \Oxy{} and \Fe{} multiplied by $E^{2.7}$, and nuclear ratios
  \CFe{} and \OFe{} as function of kinetic energy per nucleon.
  The solid lines indicate the model calculations. The contributions arising from the
  nearby source (short-dashed lines) and from the Galactic ensemble (long-dashed lines) are shown.
  The data are from ATIC-2 \citep{Panov2013,Panov2014}, CREAM \citep{Ahn2009}, CRN \citep{Muller1991}, and TRACER \citep{Ave2008,Obermeier2011}. 
  The TRACER data on nuclear ratios are those obtained in Ref.\citep{Panov2014}.
  Standard model predictions are also shown for the \CFe{} and \OFe{} ratios (dotted lines).
}\label{Fig::ccHeavyNucleiFlux}
\end{figure}
%%%%%%%%%%%%%%%%%%%%%%%%%%%%%%%%%%%%%%%%%%%%%%%%%%%%%%%%%%%%%%%%%%%%%

%%%%%%%%%%%%%%%%%%%%%%%%%
\MyTitle{Results} --- %%%
%%%%%%%%%%%%%%%%%%%%%%%%%
The model predictions are shown in Fig.\,\ref{Fig::ccHeavyNucleiFlux}
for the \OFe{} and \CFe{} ratios and for the spectra of \Fe{} and \Oxy.
The two flux components are shown as dashed lines.
The \C{} spectrum, not shown, is very similar to the \Oxy{} spectrum.
Owing spallation, the \Fe{} spectrum of the Galactic SNR population 
is slightly reshaped due to a ``missing flux'' from distant SNRs that \emph{do not} contributed to its total flux at Earth.
This effect is maximized in the primary/primary ratios \CFe{} and \OFe{} that, as seen in the figure, experience a 
remarkable spectral upturn above a few tenths of GeV/nucleon energies. 
The ATIC-2 data are described very well by the model.
These features are also present in the TRACER data, as shown in \citet{Panov2014}, but not in the CREAM data.
For the two ratios the upturn is similar and, in fact, the \CO{} ratio is featureless.
At $E\sim$\,1--10 GeV/nucleon, the flux is entirely dominated by the nearby component. 
At energies above $\sim$\,TeV/nucleon, the spallation effect
vanishes and the ratios become asymptotically representative of the spectral properties of the Galactic ensemble.
For reference, the \CFe{} and \OFe{} ratios arising from standard calculations, \ie, using one class of sources,
are plotted as dotted lines. As discussed, conventional models are unable to describe any spectral change on these ratios.
From this mechanism, similar features are expected for other ratios such as \NeFe, \MgFe, or \ArFe. 
Interestingly, an upturn in the \ArFe{} and \CaFe{} ratios was observed by HEAO3 \citep{Binns1988}.
However, these elements require more refined elaborations due to the presence of secondary components in their flux. 
Also the effective approach used here, only suitable for primary/primary nuclear ratios, suffers from several limitations.
For instance, the dilution factor $h/L$ used to estimated the average interaction rate is 
probably a too crude simplification that leaves an uncertainties on the absolute scale of $\lambda^{\rm sp}$. 
In fact, since the observed CRs are both injected/detected from/in the disk, it is improbable that during their past history
they have spent much time in the halo. 
Furthermore, due to possible inhomogeneous diffusion or convection processes, the CR transport may be either 
more confined in the disk or swept out in the halo, respectively \citep{Taillet2003,Jones1978,Tomassetti2012}.
A rigorous treatment of the problem has to account for all these unknowns which, however,  
require the use of better quality CR data. 
In particular, the \Fe{} spectrum deserves more experimental investigation.
%
%Luckily enough, 
Fortunately, we are in a proficient era for CR physics.
Ongoing and planned space experiments such as
\AMS, ISS-CREAM \citep{Seo2014}, or CALET \citep{Adriani2014Calet} will measure
these elements over a large energy range. 

%%%%%%%%%%%%%%%%%%%%%%%%%%%%%%%%%%%%%%
\MyTitle{On the nearby source} --- %%%
%%%%%%%%%%%%%%%%%%%%%%%%%%%%%%%%%%%%%%
%
In this interpretation of the \CFe{} and \OFe{} ratios, the presence of a source placed near the Solar System is a key ingredient. 
From the model presented here, such a source is identified as a local SNR ($d\sim$\,few\,100\,pc) with low magnetization ($B\sim$\,$\mu$G), 
slow shock speed ($u_{1}\sim$\,5$\times$\,10$^{7}$\,cm\,s$^{-1}$), 
and a high gas density in comparison to that of the ISM $n\sim$\,1\,cm$^{-3}$.
The accelerated spectra are rather steep ($\nu\sim$\,2.7) and limited to a maximum rigidity $R^{\rm max}\sim$\,TV.
These properties are appropriate for \emph{old} SNRs of type Ia.
Remnants of this type may be not be detectable in $\gamma$--rays any longer,
unless the rate of \p-\p{} collisions is enhanced by the presence of denser media such as molecular clouds.
In this case the emission might be sufficiently high to be detected by the \Fermi{} observatory. 
Possible examples are SNRs W44, W28, W51C or IC-443 \citep{Abdo2010W44,Abdo2010IC443,Abdo2010W28,Abdo2009W51C}. 
Indications of nearby sources in the CR spectrum are found in several recent studies
\citep{Malkov2012,Vladimirov2012,Moskalenko2003,Kachelriess2015,Nierstenhoefer2015}, 
and noticed from independent studies in connection with the local bubble \citep{Benitez2002,Moskalenko2003,ErlykinWolfendale2011}. 
Concerning the secondary production mechanism, antiprotons are also emitted from such a SNR but, similarly to the case 
of \Li-\Be-\B{} nuclei \citep{TomassettiDonato2015}, no striking signatures are expected after accounting for both source components.
In fact, the \pbarp{} ratio ``excess'' predicted in related studies \citep{BlasiSerpico2009,Kohri2015,MertschSarkar2014}
arises from one-component scenarios. 
The diffuse $\gamma$-ray emission can be used to test models involving nearby sources.
Roughly, one may expect a diffuse spectrum which is harder than that predicted by standard models, at least in the Galactic plane
where the emission is dominated by the $\pi^{0}$ production from \p-\p{} collisions.
However, the appearance of individual sources in the CR spectrum demands a different calculation scheme,
possibly beyond the usual steady-state description \citep{BlasiAmato2012,Bernard2013,Miyake2015}.
This description might be, in fact, an oversimplification of reality which does not reflect the 
\emph{stochastic nature} of SNR events and their influence on the surrounding CR flux.
Interestingly, indications of CR flux variations in the Galaxy are found by recent \Fermi{} observations of the 
diffuse $\gamma$-ray emission \citep{Ackermann2012}, and were also noted in previous studies \citep{Digel2001,Buesching2005}. 
The \Fermi{} data show harder $\gamma$-ray spectra in the inner Galaxy which are at odds with standard calculations.
While these observations might be explained in terms of SNR properties \citep{Buesching2005,Ackermann2012},
a complete calculation --possibly accounting for their space/time discreteness-- has never been attempted for $\gamma$-rays.

%%%%%%%%%%%%%%%%%%%%%%%%%%%%%
\MyTitle{Conclusions} --- %%%
%%%%%%%%%%%%%%%%%%%%%%%%%%%%%
%
This work is motivated by the search of a comprehensive, agreeable model for Galactic CRs that 
is able to account for the several puzzling features recently observed in their spectrum. 
Without the presence of nearby sources, it is difficult to interpret the ATIC-2 data in the context of known models of CR propagation.
The only alternative interpretation of the spectral upturn is the one proposed by the ATIC-2 Collaboration, 
the \emph{closed Galaxy model with bubbles} \citep{Panov2014}.
In my opinion, their model may represent a viable solution of the puzzle, but it suffers from some problems. 
In particular, it predicts a too weak rise of the elemental ratios, while requiring too steep source spectra and a too flat \BC{} ratio. 
On the contrary, the interpretation presented here accounts well for the basic observations on 
primary spectra and secondary/primary ratios. 
Despite the approximate calculation method employed in this work, it is remarkable that the spectral upturn arises 
under a scenario that is able to simultaneously account for
important features in the CR spectrum, namely, the rise in the positron fraction and the 
spectral hardening of proton and nuclei.

{\footnotesize%
  I am grateful to D. Maurin and L. Derome for discussions, and to A. Panov for kindly providing me with the ATIC-2 data.
  Other data are extracted from the \texttt{CRDB} database \citep{Maurin2014}. 
  This work is supported by the ANR LabEx grant \textsf{ENIGMASS} at CNRS/IN2P3.
}

%%%%%%%%%%%%%%%%%%%%%%%%%%%%%%%%

\end{document}